\documentclass[
 reprint,
 nofootinbib,
 amsmath,amssymb,
 aps,
 prl
]{revtex4-2}

\usepackage{graphicx}
\usepackage{hyperref}
\usepackage{comment}
\usepackage{xcolor}
\newcommand{\orcid}[1]{\href{https://orcid.org/#1}{\includegraphics[width=8pt]{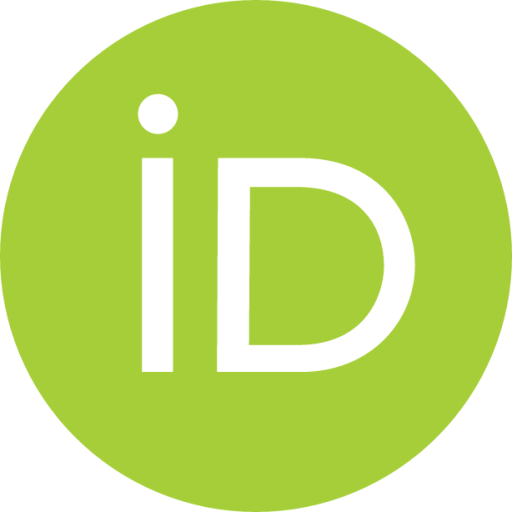}}}
\usepackage{tikz}
\usetikzlibrary{fadings, shadings, shapes, arrows.meta, positioning, calc}

\begin{document}

\title{Cherenkov Bursts are Memory Steps}

\author{Jann Zosso\orcid{0000-0002-2671-7531}}
\email{jann.zosso@uzh.ch}
\affiliation{Center of Gravity, Niels Bohr Institute, Blegdamsvej 17, 2100 Copenhagen, Denmark}
\affiliation{Physik-Institut, University of Z\"urich,
Winterthurerstrasse 190, Z\"urich, 8057, Switzerland}

\author{Guangzi Xu\orcid{0009-0002-1276-9012}}
\affiliation{Center of Gravity, Niels Bohr Institute, Blegdamsvej 17, 2100 Copenhagen, Denmark}

\author{Vitor Cardoso\orcid{0000-0003-0553-0433}}
\affiliation{Center of Gravity, Niels Bohr Institute, Blegdamsvej 17, 2100 Copenhagen, Denmark}

\date{\today}

\begin{abstract}
We show that the familiar Cherenkov burst from a finite superluminal charge track can be understood as a pair of interfering electromagnetic memory transitions. Endpoint accelerations of the track produce opposite memory steps in the transverse vector potential, such that the corresponding finite-frequency radiation is universally fixed, below the endpoint-acceleration-scale, by the memory amplitude and the time for which the temporary memory offset persists. As the Cherenkov direction is approached, the memory amplitudes grow while their separation in observer time shrinks, until the two vector-potential steps overlap and their divergent contributions cancel into a finite Cherenkov burst. This perspective opens the door to using the Askaryan effect as a setting in which individual macroscopic memory transitions may be resolved, providing a new experimental arena for memory-linked infrared physics.
\end{abstract}

\maketitle

\section{Introduction}\label{sec:Intro}

Gravitational memory \cite{Zeldovich:1974gvh,Christodoulou:1991cr,Blanchet:1992br,Thorne:1992sdb,Strominger:2014pwa} is a non-oscillatory component of gravitational radiation that leaves behind a permanent distortion of spacetime, representing a sharp time-domain manifestation of soft theorems and asymptotic symmetries. An analog phenomenon exists in electromagnetism, where the non-oscillatory component of radiation captures the Weinberg soft-photon factor \cite{Bieri:2013hqa,Weinberg_PhysRev.140.B516,He:2014cra,Campiglia:2015qka,Pasterski:2015zua,Kapec:2015ena,Garfinkle:2022dnm}. For an observer at large radius, it appears as a net offset in the transverse vector potential that results in a non-zero integral of the radiative electric field,
\begin{equation}
    \Delta \mathbf A^{\rm T}
    =
    -\int du\,\mathbf E^{\rm T}(u).
    \label{eq:memory-def}
\end{equation}
In contrast to the gravitational case whose physical effect is a net displacement of test-masses, an idealized electromagnetic memory offset results in a non-zero velocity kick of asymptotic test-charges \cite{Bieri:2013hqa}.
In practice, however, memory searches in both cases target the finite-frequency imprint of the memory transition as an observable signature of the ideal DC offset \cite{Bieri:2023btq,Favata:2010zu,Lasky:2016knh,Grant:2022bla,Inchauspe:2024ibs,Zosso:2026czc,Agazie:2025oug}.

Recent work has shown that memory effects can be parametrically enhanced when Poincar\'e invariance is effectively broken by a reduced propagation speed of the radiative field \cite{Heisenberg:2025tfh,Heisenberg:2025roe,Zosso:2025orw}, an enhancement controlled by the same denominator that appears in Cherenkov radiation \cite{Cherenkov:1934,Frank:1937,Tamm:1939}. The purpose of this work is to make the physical meaning of this denominator transparent and to present important implications for experimental searches. 

We first review the electromagnetic memory produced by the emission of a single charge in a medium with reduced phase velocity. We focus on the scale-separated regime, in which the microscopic acceleration duration and source size are unresolved, and clarify that the enhancement towards the Cherenkov direction is controlled by the map from source time to asymptotic retarded time. We then argue that the singular behavior at the critical Cherenkov direction is resolved by the physical assumption that the uniform superluminal motion of a free charge must be finite. Furthermore, we show that a pair of opposite-sign memory transitions associated with an onset and termination event is strictly equivalent, at the level of the prescribed classical current, to the standard endpoint description of a finite Cherenkov track \cite{Zas:1991ac,James:2011db,Huege:2011ut,AlvarezMuniz:2010aq}. While the strict zero-frequency memory cancels, the observationally relevant finite-frequency radiation is entirely characterized by the memory offset and by the duration for which the memory plateau persists. This description of Cherenkov bursts as radiation from interfering memory steps provides a new conceptual interpretation of their origin and finiteness condition, and identifies macroscopic Askaryan radiation as a concrete experimental probe of finite-frequency electromagnetic memory transitions \cite{Askaryan:1962,Saltzberg:2000bk,Gorham:2004ny,Miocinovic:2006it,ANITA:2006nif,AlvarezMuniz:2010aq,Bechtol:2021tyd}.

\section{Single memory steps}\label{sec:SingleStep}

We assume that electromagnetic radiation propagates with phase velocity \(v_{\rm em}\), which may be smaller than the vacuum speed of light. The results are independent of the effective structure---such as a dielectric medium or a waveguide---realizing this reduction. For simplicity, we consider a homogeneous and lossless medium large enough to work effectively in an asymptotic radiation zone, with \(v_{\rm em}\) independent of frequency over the range of interest.

In this context, consider a charge \(q\) emitted from a localized source and accelerated from rest to a constant velocity \(v_q\hat{\mathbf z}\). At frequencies below the microscopic emission scale, the process is well modeled as an instantaneous acceleration \cite{PhysRev.76.365,Jackson:1998nia,PhysRevD.68.084011}. We use coordinates with the acceleration event taken to lie at the spatial origin and define the radiation-zone retarded time as \(u\equiv t-r/v_{\rm em}\). The retarded source time \(t_r\) selected by the Green function is mapped to this asymptotic observer time as
\begin{equation}
    u\simeq U(t_r,\theta)
    \equiv
    t_r-\frac{\hat{\mathbf n}\cdot\mathbf x_q(t_r)}{v_{\rm em}}=t_r-\frac{x_q(t_r)}{v_{\rm em}}\cos\theta ,
    \label{eq:retarded-map}
\end{equation}
where \(\theta\) defines the angle between the observation direction \(\hat{\mathbf n}\) and the charge trajectory.
On the final uniform branch \(\mathbf x_q(t_r)=v_q\hat{\mathbf z}(t_r-t_1)\) after the emission at $t_1$, the associated Jacobian reads
\begin{equation}
    J(\theta)
    \equiv
    \frac{dU}{dt_r}
    =
    1-\frac{v_q}{v_{\rm em}}\cos\theta .
    \label{eq:J}
\end{equation}

\begin{figure*}[t]
\centering
\begin{minipage}{0.32\linewidth}
\centering
\includegraphics[width=\linewidth]{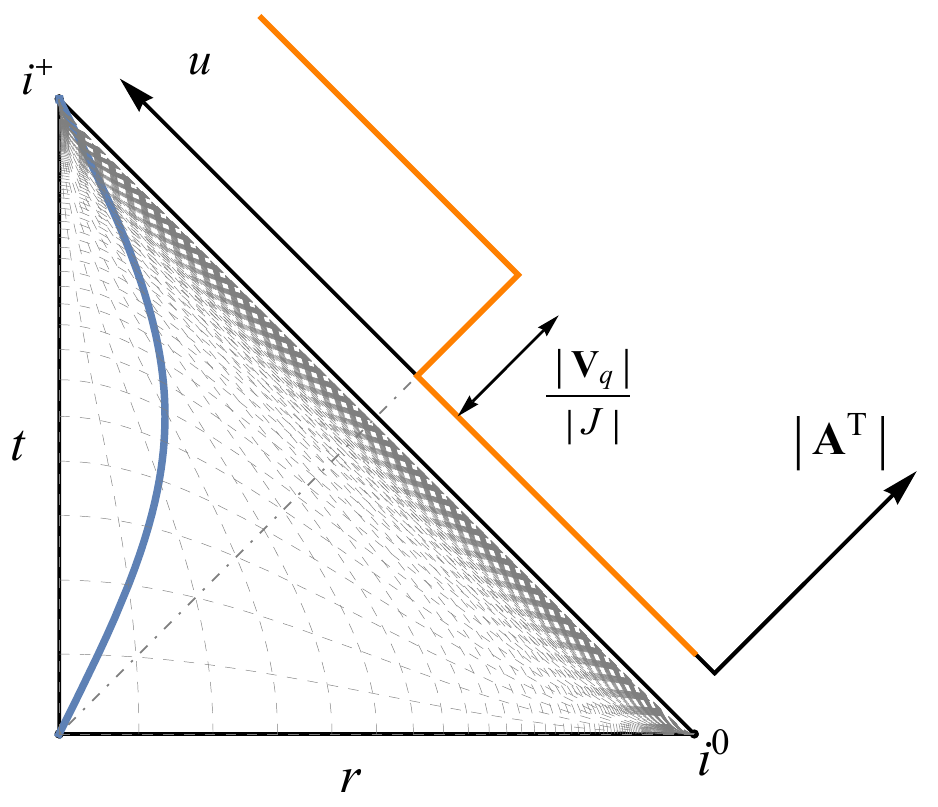}\\[-2pt]
\small (a) \(J=0.5>0\)
\end{minipage}\hfill
\begin{minipage}{0.32\linewidth}
\centering
\includegraphics[width=\linewidth]{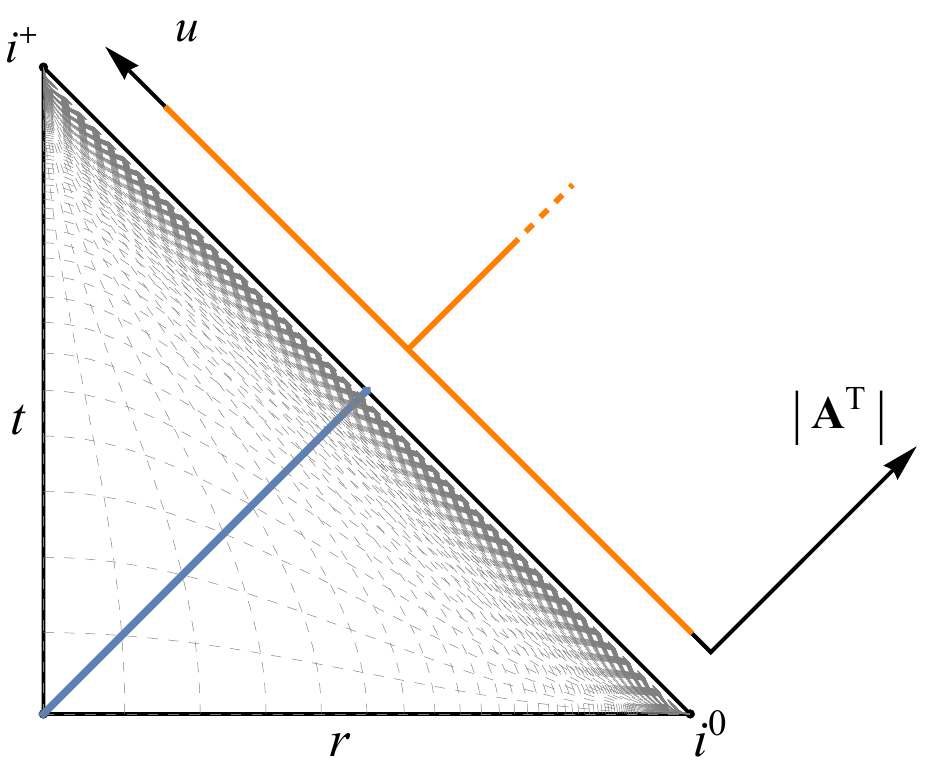}\\[-2pt]
\small (b) \(J=0\)
\end{minipage}\hfill
\begin{minipage}{0.32\linewidth}
\centering
\includegraphics[width=\linewidth]{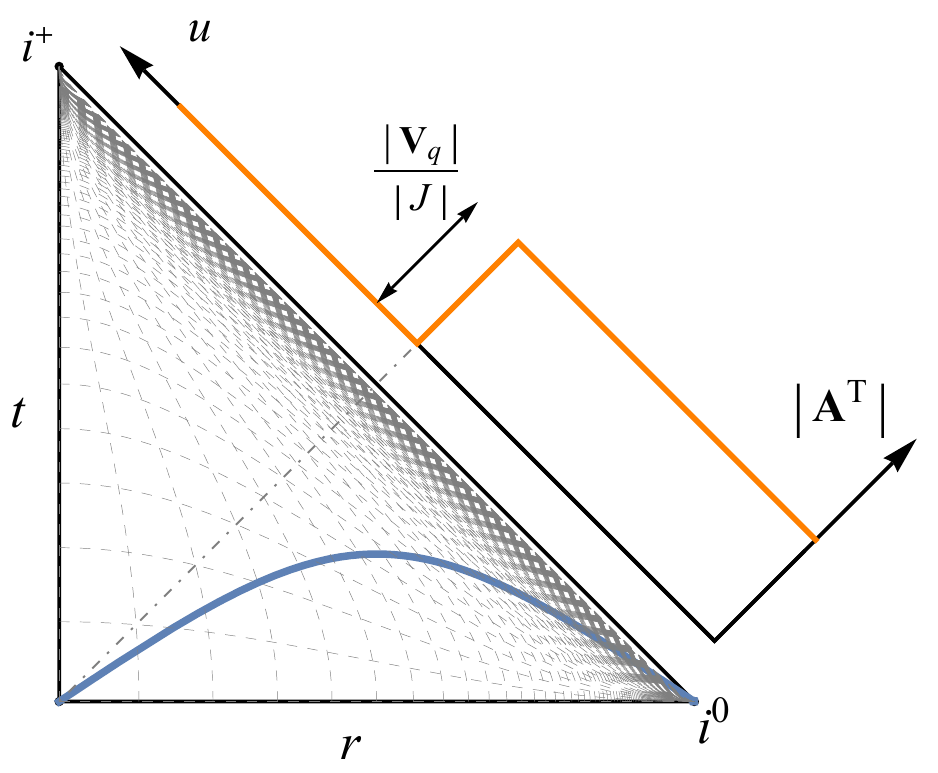}\\[-2pt]
\small (c) \(J=-0.5<0\)
\end{minipage}
\caption{\small Observer-adapted Penrose representation of an isolated onset at \(t_1=u_1=0\) followed by uniform propagation, in units with \(v_{\rm em}=1\). For each observation direction \(\hat{\mathbf n}\), the charge trajectory (blue) is projected onto the observer's radial line in the shown \((t,r)\) plane. (a) For \(J>0\), the onset produces a regular memory step. (b) At \(J=0\), the projected worldline is parallel to the outgoing effective null direction, such that the indefinite uniform propagation is compressed into a single retarded time, producing the pole. (c) For \(J<0\), the projected motion is superluminal and the emission event produces a finite memory step of opposite sign.}
\label{Fig:Penrose1}
\end{figure*}

Solving the sourced dynamical equations for the transverse vector potential in Lorenz gauge for the emitted charge described above gives a memory step with instantaneous-onset solutions \cite{Zosso:2025orw}
\begin{equation}
\mathbf A^{\rm T}(u,\theta)=
\begin{cases}
\displaystyle \frac{\mathbf V_q}{J}\,\Theta(u-u_1), & J>0,\\[5pt]
\displaystyle \frac{\mathbf V_q}{|J|}\,\Theta(u_1-u), & J<0.
\end{cases}
\label{eq:onset-solutions}
\end{equation}
Here we have defined 
\begin{equation}
     \mathbf V_q
    \equiv
    \frac{\mu q}{4\pi r}\,
    \mathbf P_{\hat n}(v_q\hat{\mathbf z}) ,
    \qquad
    \mathbf P_{\hat n}^{ij}\equiv\delta^{ij}-\hat n^i\hat n^j,
    \label{eq:J-and-V}
\end{equation}
with \(\mu\) the magnetic permeability of the medium and \(u_1\equiv U(t_1,\theta)\). The associated memory offset as defined in Eq.~\eqref{eq:memory-def} is therefore
\begin{equation}
    \Delta\mathbf A^{\rm T}
    =
    \frac{\mathbf V_q}{J},\qquad J\neq0.
    \label{eq:onset-memory}
\end{equation}
The \(J>0\) case corresponds to an ordinary electromagnetic memory step. If \(v_q>v_{\rm em}\) there exists a second regime satisfying \(J<0\), for which the uniform trajectory is mapped to an earlier observer time resulting in a sign flip of the memory amplitude. The transition between the two regimes occurs at
\begin{equation}
    J=0
    \quad\Longleftrightarrow\quad
    \cos\theta_c=\frac{v_{\rm em}}{v_q},
    \label{eq:cherenkov-angle}
\end{equation}
which is the Cherenkov angle that manifests as a pole in the onset memory. The corresponding causal structure is most transparently represented by the observer-adapted \(1+1\)-dimensional Penrose diagrams in Fig.~\ref{Fig:Penrose1}, where the charge trajectory is projected onto the corresponding line of sight. In particular, the critical case obtains a direct geometric interpretation: the projected worldline becomes parallel to the effective null direction, such that an indefinitely long uniform motion is compressed into a single asymptotic observer time $u$.

\section{Memory steps from finite tracks}\label{sec:FiniteTrack}

The divergence of the isolated superluminal onset at J=0 is tied to the assumption that the charge continues in uniform superluminal motion forever. This is because maintaining exact uniform superluminal motion indefinitely would require an indefinite external energy supply, since the charge is losing energy in the form of Cherenkov radiation. A prescribed infinite uniform current is a legitimate idealized source for Maxwell's equations, but it does not describe the self-consistent motion of a free particle once cumulative energy-loss into Cherenkov modes becomes important. Determining the true trajectory of a freely propagating superluminal charge in such a regime would require solving the coupled radiation-reaction problem.\footnote{The formal long-track limit should therefore be understood as a useful phase-coherence limit of the field equations, rather than as the literal infinite-time trajectory of a free particle.} 


\begin{figure*}[t]
\centering
\begin{minipage}{0.32\linewidth}
\centering
\includegraphics[width=\linewidth]{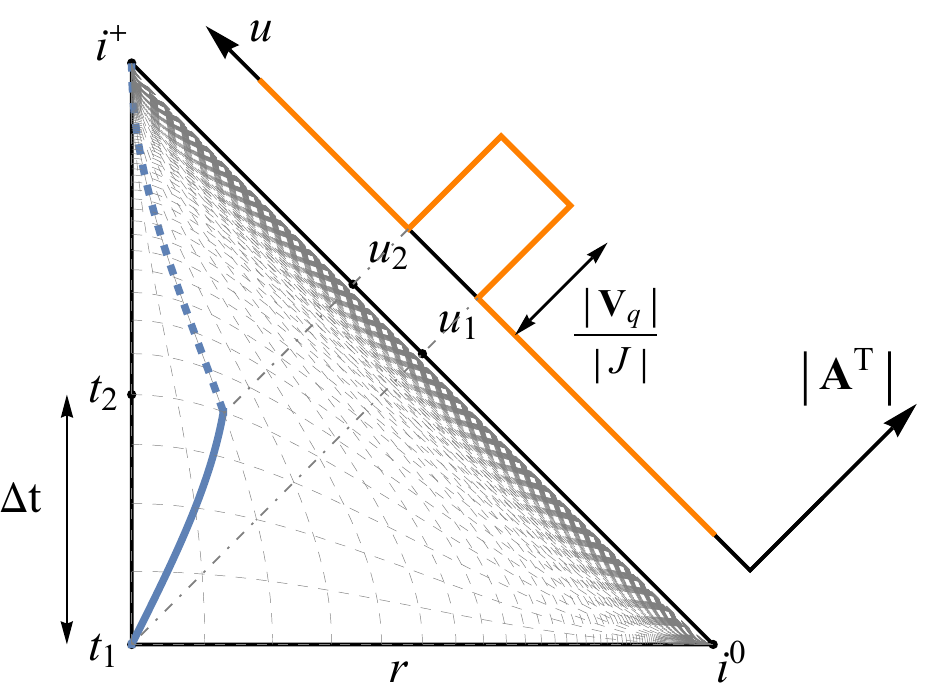}\\[-2pt]
\small (a) \(J=0.5>0\)
\end{minipage}\hfill
\begin{minipage}{0.32\linewidth}
\centering
\includegraphics[width=\linewidth]{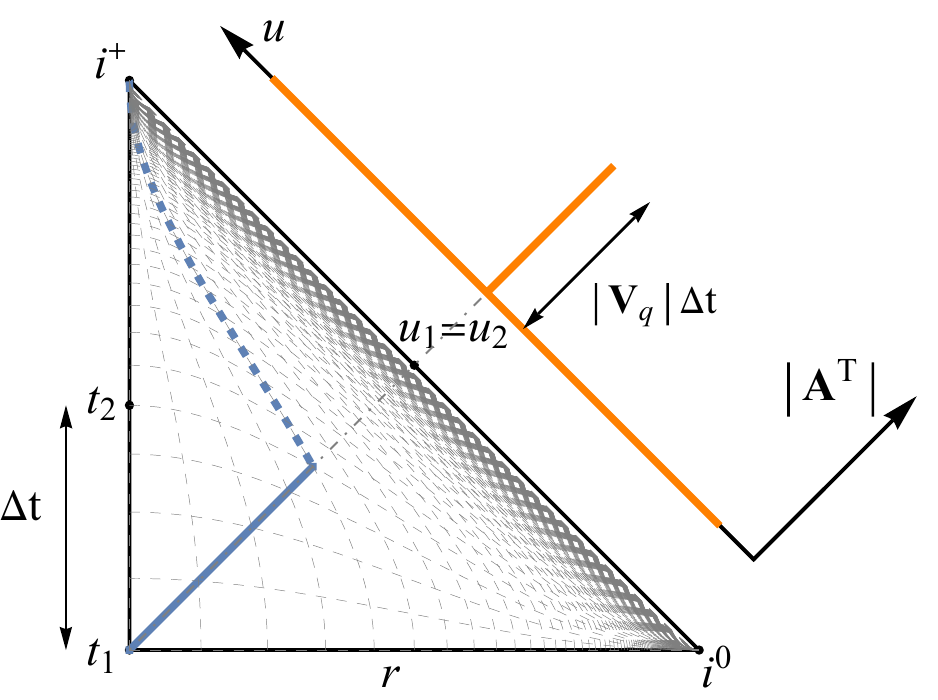}\\[-2pt]
\small (b) \(J=0\)
\end{minipage}\hfill
\begin{minipage}{0.32\linewidth}
\centering
\includegraphics[width=\linewidth]{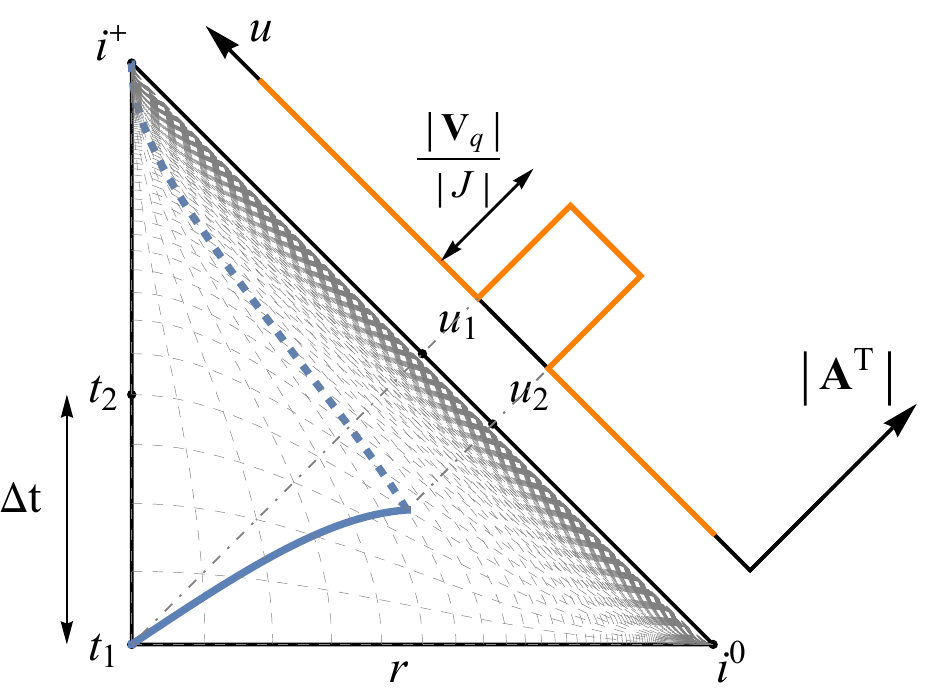}\\[-2pt]
\small (c) \(J=-0.5<0\)
\end{minipage}
\caption{\small Observer-adapted Penrose diagrams for a finite track with constant projected velocity \(v_q\cos\theta\), using the same construction as Fig.~\ref{Fig:Penrose1}. The endpoints produce opposite memory steps at \(u_1\) and \(u_2\), separated by \(\Delta u=J(\theta)\Delta t\). (a) For \(J>0\), the onset reaches the observer before the termination pulse. (b) At \(J=0\), \(u_1=u_2\) and the two divergent endpoint steps overlap into the finite Cherenkov burst, with vector-potential weight \(|\mathbf V_q|\Delta t\). (c) For \(J<0\), the ordering of the memory steps is reversed from the perspective of the asymptotic observer. }
\label{Fig:Penrose2}
\end{figure*}

In this work, we will not address such a non-linear problem, but instead model a finite superluminal motion by considering a charge that is instantaneously accelerated from rest at source time \(t_1\), moves with constant velocity \(v_q\hat{\mathbf z}\), and is instantaneously decelerated again at \(t_2\). In a regime in which the dominant velocity changes are localized scattering, creation, or absorption events, and radiative backreaction is negligible during the intermediate motion, such a finite track is the physically relevant description.
Defining
\begin{equation}
    \Delta t\equiv t_2-t_1,
    \qquad
    u_i\equiv U(t_i,\theta).
    \label{eq:finite-track-times}
\end{equation}
we have to leading order in \(1/r\),
\begin{equation}
    \Delta u\equiv u_2-u_1=J(\theta)\Delta t.
    \label{eq:UdiffToT}
\end{equation}

The associated radiative transverse vector potential in the instantaneous-acceleration approximation analogous to Eq.~\eqref{eq:onset-solutions} is 
\begin{equation}
    \mathbf A_{\Delta t}^{\rm T}(u,\theta)
    =
    \frac{\mathbf V_q}{|J|}
    \left[\Theta(u-u_<)-\Theta(u-u_>)\right],
    \label{eq:finite-track-rectangle}
\end{equation}
where \(u_<\equiv\min(u_1,u_2)\) and \(u_>\equiv\max(u_1,u_2)\). This is a finite memory plateau with vanishing net memory for the complete track, \(\Delta\mathbf A_{\Delta t}^{\rm T}=0\), but each endpoint separately carries a memory step of amplitude \(|\Delta\mathbf A^{\rm T}|=|\mathbf V_q|/|J|\). The corresponding radiative electric field,
\begin{equation}
    \mathbf E_{\Delta t}^{\rm T}(u,\theta)
    =
    \frac{\mathbf V_q}{|J|}
    \left[\delta(u-u_>)-\delta(u-u_<)\right],
    \label{eq:finite-track-E}
\end{equation}
therefore consists of two pulses of non-zero and opposite area, \(\mp |\Delta\mathbf A^{\rm T}|\). These individual pulses precisely correspond to the characteristic finite-time signature of electromagnetic memory \cite{Bieri:2013hqa,Bieri:2023btq,Zosso:2025orw}.


Most importantly, the finite-track vector potential no longer exhibits a pole at \(J\rightarrow 0\), hence $\theta\rightarrow\theta_c$, because while the endpoint amplitudes still grow as \(1/|J|\), the observer-time separation of the two steps shrinks as \(|\Delta u|=|J|{\Delta t}\). Concretely, equations~\eqref{eq:finite-track-rectangle} and \eqref{eq:finite-track-E} have the well-defined distributional limit
\begin{align}
   \lim_{J\to0}\mathbf A_{\Delta t}^{\rm T}(u,\theta)
    &=\mathbf V_q\,{\Delta t}\,\delta(u-u_1),\label{eq:finite-track-cherenkov-time-domainA}\\
    \lim_{J\to0}\mathbf E_{\Delta t}^{\rm T}(u,\theta)
    &=-\mathbf V_q\,{\Delta t}\,\delta'(u-u_1),
    \label{eq:finite-track-cherenkov-time-domain}
\end{align}
where the finite track is compressed onto the single critical observer time \(u_1=u_2\). 
The remaining distributional singularity is the ultraviolet artifact of the instantaneous-acceleration approximation: finite endpoint duration replaces it by a finite pulse of some timescale $\tau$. All expressions in this and the next section are universal as long as this effective microscopic endpoint scale \(\tau\) remains unresolved. The cancellation of the \(1/J\) endpoint poles, by contrast, follows from the finite duration of the track. An explicit distributional derivation and an effective smoothing description are given in the End Matter.

Thus, along the critical Cherenkov direction the individual memory steps no longer exist as separately resolved pulses. The electric-field area of this single pulse vanishes, so no residual DC memory remains, while the overlap leaves a finite Cherenkov burst. Figure~\ref{Fig:Penrose2} makes the causal origin of this cancellation explicit.

\section{Universal finite-frequency regime}\label{sec:UniversalBand}

That the onset--termination computation in the scale-separated regime above is entirely characterized by the memory amplitude of the individual steps and by the duration for which the temporary memory offset persists is most cleanly seen in the Fourier domain. With \(\widetilde{\mathbf E}^{\rm T}(\omega)=\int du\,e^{i\omega u}\mathbf E^{\rm T}(u)\), Eq.~\eqref{eq:finite-track-E} gives 
\begin{equation} 
\widetilde{\mathbf E}^{\rm T}_{\Delta t}(\omega,\theta)=\Delta\mathbf A^{\rm T}e^{i\omega u_1}\left(e^{i\omega\Delta u}-1\right), 
\label{eq:finite-track-spectrum-memory} 
\end{equation} 
so the amplitude is fixed by the memory offset and the phase difference by the observer-time duration. Equivalently, 
\begin{equation} 
\widetilde{\mathbf E}_{\Delta t}^{\rm T}(\omega,\theta)=i\omega\,\mathbf V_q\Delta t\,e^{i\omega\left(u_1+J\Delta t/2\right)}{\rm sinc}\!\left(\frac{\omega J\Delta t}{2}\right), 
\label{eq:finite-track-spectrum-sinc} 
\end{equation} 
with \({\rm sinc}(x)=\sin x/x\). It is now manifest how the shrinking observer-time duration compensates the divergent memory amplitude, leading to the finite Cherenkov limit 
\begin{equation}\label{eq:CherenkovE-Fourier} 
\lim_{J\rightarrow 0}\widetilde{\mathbf E}_{\Delta t}^{\rm T}(\omega,\theta)=\widetilde{\mathbf E}_{\Delta t}^{\rm T}(\omega,\theta_c)=i\omega\mathbf V_q\Delta t\,e^{i\omega u_1}. 
\end{equation}

This linear frequency scaling is, however, not intrinsically Cherenkov: it occurs at any direction of the finite-track radiation for frequencies below which the two opposite memory steps remain unresolved \(\omega|\Delta u|\ll1\). Indeed, in this limit the signal in Eq.~\eqref{eq:finite-track-spectrum-sinc} reduces to \(\widetilde{\mathbf E}_{\Delta t}^{\rm T}=i\omega\mathbf V_q\,{\Delta t}\,e^{i\omega u_1}+O(\omega^2\Delta u^2)\) which to leading order coincides with Eq.~\eqref{eq:CherenkovE-Fourier}. What is special near the Cherenkov direction is that \(\Delta u=J\Delta t\) becomes small, so the unresolved-memory condition remains valid to progressively higher frequencies. Conversely, the individual memory steps become resolvable above the characteristic scale \(\omega_{\rm mem}\sim|\Delta u|^{-1}=|(J|\Delta t)^{-1}\), allowing for a genuine memory-resolving finite-frequency window
\begin{equation}
    \omega_{\rm mem}
    \lesssim\omega\ll\tau^{-1},
    \label{eq:universal-window-main}
\end{equation}
where \(\tau\) is the observed microscopic scale of an individual transition introduced in the previous section.

The radiation from emission and absorption of a superluminal charge presented here is exactly equivalent, at the level of the prescribed current, to the standard endpoint and Zas--Halzen--Stanev (ZHS) finite-track descriptions of Cherenkov bursts \cite{Zas:1991ac,James:2011db,Huege:2011ut,AlvarezMuniz:2010aq}. Yet, the interpretation is novel: in the time domain the burst is identified as a pair of opposite memory transitions, while in the frequency domain the total radiation is their interference. This viewpoint makes transparent why a finite superluminal track is special near the Cherenkov direction: the retarded-time map compresses the otherwise resolvable memory transitions into a coherent finite-band burst, a mechanism that is absent in subluminal memory radiation emission. Away from the cone, in the window of Eq.~\eqref{eq:universal-window-main}, the two memory transitions can instead be resolved through their individual non-zero electric-field areas.

\section{Askaryan radiation as finite-frequency memory}\label{sec:ExperimentalEv}

We now show that the understanding above in particular implies that there is a direct correspondence between memory transitions and the Askaryan effect in dense media \cite{Askaryan:1962,Saltzberg:2000bk,Gorham:2004ny,Miocinovic:2006it,ANITA:2006nif,AlvarezMuniz:2010aq,Bechtol:2021tyd}, making associated observations a potential experimental probe for memory-related theory. We provide a short description of Askaryan radiation and details of the arguments in the End Matter.

\begin{figure}[h!]
    \centering
    \begin{tikzfadingfrompicture}[name=smoothfade]
        \shade[shading=radial, inner color=transparent!0, outer color=transparent!100] (2.9,0) circle (.95cm);
    \end{tikzfadingfrompicture}

    \begin{tikzpicture}[scale=1.2, every node/.style={scale=1}]
        
        
        
        


        
        


        \draw[thick,->] (-0.8,0) -- (5.2,0) node[right] {$z$};
        
        \fill[blue!40!white, opacity=0.4]
        (0.9,0)
        .. controls (1.3,0) and (1.4,-0.72) .. (1.8,-0.72)
        -- (2.80,-0.72)
        .. controls (3.2,-0.72) and (3.3,0) .. (3.7,0)
        -- cycle;
        \draw[blue!70!black, thick]
        (0.9,0)
        .. controls (1.3,0) and (1.4,-0.72) .. (1.8,-0.72)
        -- (2.80,-0.72)
        .. controls (3.2,-0.72) and (3.3,0) .. (3.7,0);
        \node[blue!70!black] at (2.3,-0.45) {$Q_{\rm eff}(t_r)$};


        \coordinate (P) at (3.15,0);
        \coordinate (N) at ($(P)+(.5,.74)$);
        \draw[orange!85!black, thick,->] (P) -- (N) node[above right] {$\hat{\mathbf{n}}$};

        \coordinate (P1) at (1.35,0);
        \path (P1) -- ++(.5,.74) coordinate (dir);
        \coordinate (N1) at ($(P1)!(N)!($(P1)+(.5,.74)$)$);
        \draw[orange!85!black, thick,->] (P1) -- (N1) ;

        \draw[dashed, orange!85!white] (N) -- (N1);

        \draw[orange!85!black, thick] (P) ++(0.5,0) arc[start angle=0, end angle=57, radius=0.5];
        \node[orange!85!black, thick] at ($(P)+(0.64,0.32)$) {$\theta_c$};

        \draw[orange!85!white, opacity=0.4] (P)++(0.89,0) arc (0:180:.89);
        \draw[orange!85!white, opacity=0.4] let \p1 = ($ (N1) - (P1) $), 
                            \n1 = {veclen(\x1,\y1)} in
        ($(P1) + (\n1, 0)$) arc (0:180:\n1);
        
        \filldraw[fill=gray!40, draw=black] (-1, -.05) rectangle ++(.6, 0.11);       
    \end{tikzpicture}
    \caption{Schematic illustration of an effective Askaryan charge profile \(Q_{\rm eff}(t_r)\) propagating along \(\hat z\). The buildup and decay memory pulses arrive simultaneously at an asymptotic observer at \(\theta_c\) (\(J=0\)), with maximal amplitude enhancement. Away from the cone the pulses arrive separately and may be resolved.}
    \label{fig:AskaryanRadiationSketch}
\end{figure}
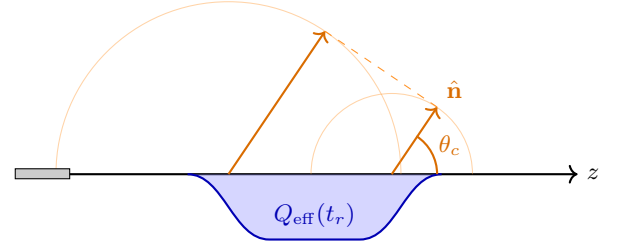

It is well known that coherent Askaryan radiation--arising from an energetic beam-like cascade following the impact of a high-energy particle in a dense medium--is well described as resulting from a temporary effective negative charge excess \(Q_{\rm eff}\) propagating with the shower front at approximately constant velocity \(v_Q>v_{\rm em}\) \cite{Gorham:2004ny,AlvarezMuniz:2010aq}. Up to form factors associated with finite shower size and detector response, and using the same conventions as before, the resulting coherent radiation-carrying vector potential is characterized by
\begin{equation}
    \mathbf A_{\rm eff}^{\rm T}(u,\theta)
    \simeq
    \frac{\mu}{4\pi r}\,
    \mathbf P_{\hat n}\mathbf v_Q
    \int dt_r\,
    Q_{\rm eff}(t_r)\,
    \delta\!\left(u-J(\theta)t_r\right),
    \label{eq:Askaryan-memory-uniform}
\end{equation}
where \(J(\theta)=1-(v_Q/v_{\rm em})\cos\theta\). For \(J\neq0\), and when the longitudinal profile is resolved in observer time,
\begin{equation}
    \mathbf A_{\rm eff}^{\rm T}(u,\theta)
    \simeq
    \frac{\mathbf V_Q(u,\theta)}{|J(\theta)|},\;\;\mathbf E_{\rm eff}^{\rm T}(u,\theta)
    \simeq -
    \frac{\partial_u\mathbf V_Q(u,\theta)}{|J(\theta)|},
    \label{eq:Askaryan-memory-effective}
\end{equation}
with
\begin{equation}
     \mathbf V_Q(u,\theta)\equiv \frac{\mu Q_{\rm eff}\!\left(t_r(u,\theta)\right)}{4\pi r} \mathbf P_{\hat n}\mathbf (v_Q\hat{\mathbf z})
\end{equation}
At \(J=0\Leftrightarrow \theta=\theta_c\), the inverse map \(t_r(u,\theta)\) is lost and the entire source-time current is compressed into the charge-profile analogue of Eq.~\eqref{eq:finite-track-cherenkov-time-domainA}, characterized by \(\int dt_r\,Q_{\rm eff}(t_r)=\Delta t_Q\langle Q_{\rm eff}\rangle\).
In Fig.~\ref{fig:AskaryanRadiationSketch} we provide a visual representation of the situation.

It follows from this effective description that Askaryan radiation represents a macroscopic realization of superluminal finite-tracks discussed above and is therefore directly related to finite-frequency memory transitions, while the Cherenkov geometry similarly controls their coherence and retarded-time compression. Again, the complete Askaryan shower has vanishing strict DC memory, but a finite-frequency memory search can target a resolved finite-time transition with non-zero electric-field area that indicates a finite change in \(\mathbf A^{\rm T}\) across an individual burst. The difference is that instead of a physical acceleration of the charge, the associated radiation is triggered directly by the time-dependence of the effective charge.  

Existing Askaryan measurements already access the key ingredients of this description. Coherent, linearly polarized radio pulses have been observed in dense media \cite{Saltzberg:2000bk,Gorham:2004ny}; broadband time-domain measurements have reconstructed the electric-field pulse together with its magnitude and phase \cite{Miocinovic:2006it}; longitudinal scans have tracked the simulated charge-excess profile \cite{Gorham:2004ny}; and controlled angular scans have mapped the radiation across the Cherenkov cone \cite{ANITA:2006nif,Bechtol:2021tyd}. These measurements were optimized primarily for the coherent Cherenkov burst, rather than for resolving the buildup and decay of the charge-excess profile as separate transitions. A dedicated phase-calibrated scan at small but non-zero \(J\) could instead target this regime directly. For illustration, a representative shower duration \(\Delta t_Q\sim3\)--\(10\,\mathrm{ns}\) and \(|J|\sim0.03\)--0.1 give a memory-resolving scale
\begin{equation}\label{eq:MemoryResolvingScale}
    f_{\rm mem}=\frac{\omega_{\rm mem}}{2\pi}\sim\frac{1}{2\pi|J|\Delta t_Q}
    \sim0.2\text{--}2\,\mathrm{GHz},
\end{equation}
overlapping established Askaryan radio bands.

These scalings provide a concrete roadmap for a dedicated memory-transition measurement. The central trade-off is between the Cherenkov enhancement at small \(|J|\) and the increasing observer-time separation of the transitions at larger \(|J|\). An optimized experiment should therefore target the smallest non-zero \(|J|\) for which \(\omega_{\rm mem}\) remains within a phase-calibrated bandwidth, while favoring an extended charge-excess profile with sufficiently sharp rise and fall, \(\tau\ll |J|\Delta t_Q\), such that a genuine memory plateau is resolved within the window of Eq.~\eqref{eq:universal-window-main}. The corresponding source, geometry, and detector optimization is discussed in the End Matter. Such a measurement would provide a controlled finite-frequency probe of electromagnetic memory transitions and a testing ground for their connection to the soft-photon Ward identity and large-\(U(1)\) charge symmetry through the universal source and angular dependence \cite{Weinberg_PhysRev.140.B516,He:2014cra,Campiglia:2015qka,Pasterski:2015zua,Kapec:2015ena}.

\section{Conclusion}\label{sec:Conclusion}

We have shown that finite-track Cherenkov bursts admit a time-domain description in terms of electromagnetic memory transitions, providing a physical resolution of the divergent isolated-onset limit identified in previous work \cite{Zosso:2025orw}. This connection identifies media and slow-wave systems as new arenas in which memory-linked infrared physics can be studied. In particular, coherent Askaryan radiation provides a setting in which macroscopic memory transitions can in principle be resolved away from the exact Cherenkov direction, enabling indirect observational tests of electromagnetic memory with the potential to pave the way for controlled laboratory access to infrared structures of field theory closely related to gravitational memory.

The connection between a Poincar\'e-breaking realization of electromagnetic memory and Cherenkov bursts also raises questions for future theoretical research. Going beyond the scale-separated instant-acceleration approximation considered here requires resolving the microscopic acceleration or absorption profile. In this high-frequency transition region the retarded-time map is expected to exhibit a genuine geometric distinction between the two sides of the Cherenkov cone, including a fold caustic and additional high-frequency enhancement inside the cone \cite{Zosso:inprep}. More broadly, Cherenkov-type phase-matching phenomena arise in a wide range of settings, including radiation emission in vector-tensor theories of gravity, inertial superradiance and gravity-induced electromagnetic emission by compact objects \cite{Dokuchaev:1964,Bekenstein:1998nt,Khlebnikov:2025xra,Heisenberg:2025tfh,Heisenberg:2025roe}, suggesting that it will be interesting to investigate how far the memory perspective developed here extends beyond ordinary electromagnetic Cherenkov radiation. Finally, this viewpoint opens a route toward understanding soft theorems and asymptotic symmetries within the infrared triangle when Poincar\'e invariance is effectively broken.

\begin{acknowledgments}
We thank David Garfinkle for useful discussions. JZ is supported by funding from the Swiss National Science Foundation (Grant No. 222346 and 242481), the Janggen-P\"ohn-Foundation and was partially supported by a grant from the Simons Foundation International and the Simons Foundation (SFI-MPS-BH-00012593-01). The Center of Gravity is a Center of Excellence funded by the Danish National Research Foundation under Grant No. 184. 
This work is further funded by the European Union (ERC grant GWSky/ 101167314). Views and opinions expressed are however those of the author(s) only and do not necessarily reflect those of the European Union or the European Research Council Executive Agency. Neither the European Union nor the granting authority can be held responsible for them.
\end{acknowledgments}

\paragraph*{Data availability.} No data were created or analyzed in this study.

\bibliography{references}

\appendix

\section{Critical finite-track limit}\label{app:finiteness}

We provide here the explicit steps to show that the critical singularity of an isolated onset is absent for a complete finite track. The finite result follows from the finite duration of the uniform branch, not from smoothing the endpoints or introducing dispersion.

Observe first that after writing \(t_r=t_1+s\), with \(0\le s\le\Delta t\), the retarded-time map of the uniform branch can be written as \begin{equation} 
U(t_1+s,\theta)=u_1+J(\theta)s.
\end{equation} 
Thus, as \(J\to0\), every source time on the finite track is mapped to the same observer time, in particular \(u_1=u_2\). We now evaluate Eq.~\eqref{eq:finite-track-rectangle} in this limit. The difference of step functions in the general solution is nonzero only on the plateau interval \(u_<<u<u_>\), such that for \(f(u)\) a smooth test function we have
\begin{align} 
\int du\,\mathbf A_{\Delta t}^{\rm T}(u,\theta)f(u)&=\frac{\mathbf V_q}{|J|}\int_{u_<}^{u_>}du\,f(u)\nonumber\\ 
&=\mathbf V_q\int_0^{\Delta t}ds\,f(u_1+Js)\nonumber\\ &\xrightarrow[J\to0]{}\mathbf V_q\,\Delta t\,f(u_1).
\end{align} 
Hence, in the distributional sense, 
\begin{align} \lim_{J\to0}\mathbf A_{\Delta t}^{\rm T}(u,\theta)&=\mathbf V_q\,\Delta t\,\delta(u-u_1),\\ 
\lim_{J\to0}\mathbf E_{\Delta t}^{\rm T}(u,\theta)&=-\mathbf V_q\,\Delta t\,\delta'(u-u_1). 
\end{align} 
Here \(\delta'\) means \(\int du\,\delta'(u-u_1)f(u)=-f'(u_1)\). Thus the complete finite track has a well-defined critical limit, even though the two individual endpoint memory amplitudes diverge before the limit is taken.

The remaining distributional singularity is a consequence of the instantaneous-endpoint approximation. Indeed, if the endpoint transitions are smoothed over an effective observed time scale \(\tau\), the delta distribution above is replaced by a finite pulse profile. Concretely, one may write
\begin{align}
    \mathbf A^{\rm T}_{\Delta t}(u,\theta_c)
    &\simeq
    \mathbf V_q\,\Delta t\,S_\tau(u-u_1),\\
    \mathbf E^{\rm T}_{\Delta t}(u,\theta_c)
    &\simeq
    -\mathbf V_q\,\Delta t\,\partial_u S_\tau(u-u_1),
\end{align}
where \(S_\tau\) is a normalized localized function of characteristic width \(\tau\), whose detailed shape depends on the endpoint dynamics, source size, dispersion, and detector response. The corresponding Fourier-domain expression is
\begin{equation}
    \widetilde{\mathbf E}^{\rm T}_{\Delta t}(\omega,\theta_c)
    \simeq
    i\omega\,\mathbf V_q\,\Delta t\,
    e^{i\omega u_1}\,
    \widetilde S_\tau(\omega),
\end{equation}
with \(\widetilde S_\tau(\omega)\equiv \int du\,e^{i\omega u}S_\tau(u)\). Thus the sharp-endpoint result \(i\omega\mathbf V_q\Delta t\,e^{i\omega u_1}\) is recovered for \(\omega\tau\ll1\), where \(\widetilde S_\tau(\omega)\simeq1\). At frequencies that resolve the transition, \(\omega\tau\gtrsim1\), the form factor \(\widetilde S_\tau(\omega)\) modifies the sharp-endpoint spectrum and suppresses its ultraviolet growth. In particular, for any localized smooth profile with integrable \(\partial_uS_\tau\), the Fourier transform of \(\partial_uS_\tau\) vanishes as \(|\omega|\to\infty\), and hence so does the electric-field spectrum.


\section{Experimental optimization of Askaryan memory transitions}\label{app:Askaryan}

The Askaryan effect occurs when a high-energy particle initiates an energetic beam-like electromagnetic shower in a dielectric medium with reduced electromagnetic phase velocity, producing a net negative charge excess. Pair production and bremsstrahlung generate the forward-moving cascade, Compton and M\o ller scattering add medium electrons while positrons are removed by annihilation and absorption. In the coherent radio regime, the many microscopic emission and absorption transitions add into the macroscopic Askaryan radiation \cite{Askaryan:1962,Saltzberg:2000bk,Gorham:2004ny,Miocinovic:2006it,ANITA:2006nif,AlvarezMuniz:2010aq,Bechtol:2021tyd}.\footnote{Endpoint-formalism simulations reproduce controlled laboratory Askaryan amplitudes at the few-percent level after corrections for the measured detector response are included \cite{Bechtol:2021tyd}.} The connection to memory, however, goes well beyond endpoint formalism matching and persists at the macroscopic scale.

The radiation emission from the resulting macroscopic negative charge \(Q_{\rm eff}\), propagating with the shower front at approximately constant velocity \(v_Q\simeq c\), is illustrated in Fig.~\ref{fig:AskaryanRadiationSketch}.
In contrast to the single-particle finite-track problem the radiation is generated by the evolution of \(Q_{\rm eff}\), rather than endpoint accelerations. For \(J\neq0\), Eq.~\eqref{eq:Askaryan-memory-effective} implies that the resolved transverse vector potential is directly proportional to the effective charge development, while the electric field follows its derivative
\begin{equation}
    \mathbf E_{\rm eff}^{\rm T}(u,\theta)
    \simeq
    -\frac{\mu}{4\pi r}\,
    \frac{\mathbf P_{\hat n}\mathbf v_Q}{|J(\theta)|J(\theta)}
    \frac{dQ_{\rm eff}}{dt_r},
    \label{eq:Askaryan-effective-E-resolved}
\end{equation}
with the source time evaluated at the retarded solution.

For a dedicated memory measurement, the relevant source parameter is the physical longitudinal scale \(\Delta\ell_Q\simeq v_Q\Delta t_Q\) over which the macroscopic charge excess develops and disappears. The range \(\Delta t_Q\sim3\)--\(10\,\mathrm{ns}\) used in Eq.~\eqref{eq:MemoryResolvingScale} corresponds for \(v_Q\simeq c\) to \(\Delta\ell_Q\sim0.9\)--\(3\,\mathrm{m}\), comparable to the several-meter electromagnetic showers already realized in \cite{Saltzberg:2000bk}. Experimentally, the coherent signal strength and longitudinal duration provide partly independent optimization knobs: increasing the number of particles at fixed energy primarily increases \(Q_{\rm eff}\) and hence the signal-to-noise ratio, whereas extending the charge-excess profile can be pursued through higher energies of incident-particles and the choice of a target material with a favorable shower-development scale. The incident-particle spectrum, target material, and beam--target geometry may therefore be optimized for a large \(\Delta t_Q\). Together with sufficiently sharp buildup and decay, this lowers \(\omega_{\rm mem}\) while preserving a broad memory-resolving window, allowing the transitions to remain separately resolvable closer to the Cherenkov cone.

The angular optimization can be made explicit by expanding near the Cherenkov direction,
\begin{equation}
    J(\theta)\simeq\frac{v_Q}{v_{\rm em}}\sin\theta_c\,(\theta-\theta_c),
    \label{eq:Askaryan-J-angular}
\end{equation}
so that the representative range \(|J|\sim0.03\)--0.1 corresponds to angular offsets of order a few degrees in typical dense dielectrics. A movable receiver array can therefore scan from the strongly compressed Cherenkov pulse into the regime of separately resolved buildup and decay. Measurements at comparable \(|J|\) on both sides of the cone provide an additional consistency test, since the longitudinal profile is mapped into observer time with opposite ordering for \(J>0\) and \(J<0\). Very close to \(\theta_c\), however, the lateral shower extent introduces an additional pulse-width scale and the one-dimensional longitudinal description breaks down \cite{AlvarezMuniz:2010aq}. The minimum useful \(|J|\) is therefore set by the requirement that \(|J|\Delta t_Q\) remain larger than the effective width of an individual transition after transverse shower size, propagation, and detector response are included. Multipath propagation provides a further experimental systematic: internal target reflections may generate delayed contributions on nanosecond timescales comparable to the separations of interest here \cite{Bechtol:2021tyd}, and should therefore be minimized or independently calibrated. Resulting memory-optimized measurements could not only test the predicted \(|J|\)-dependent transition separation but also the signed electric-field areas.

\end{document}